\documentclass[a4paper,11pt]{article}
\pdfoutput=1 

\usepackage{jinstpub} 

\title{Precision Measurement of Pixel Sensor Capacitance}

\author[a,1]{H. Krüger,\note{Corresponding author.}}
\author[a]{E. Kimmerle}

\affiliation[a]{University of Bonn, Physics Department,\\Nussallee 12, 53115 Bonn, Germany}

\emailAdd{krueger@physik.uni-bonn.de}

\abstract{The capacitance of the charge collection node of a sensor system is an important parameter for the design of the analog front-end electronics. The analog front-end of high-granularity sensors like for example hybrid pixel detectors need to be optimized for timing resolution, power consumption, and electronics noise - parameters which all depend on the pixel capacitance. Current pixel detector developments for the HL-LHC upgrade typically use silicon sensors with a pixel size in the order of $50 \times \SI{50}{\square\micro\meter}$ which have a pixel capacitance of several tens of fF, depending on the sensor geometry. We have developed a dedicated integrated circuit to be bump-bonded to a pixel sensor, which enables a precise pixel capacitance measurement by using the charge-based capacitance measurement method. In this paper, we will describe the measurement method and the implementation of the capacitance measurement chip (Pixcap65) and show measurement results of a planar pixel sensor whose pixel capacitance is influenced by variations of the implant geometry.}

\keywords{Hybrid detectors, Pixelated detectors and associated VLSI electronics, Front-end electronics for detector readout, VLSI circuits}

\begin{document}
\maketitle
\flushbottom

\section{Introduction}
\label{sec:intro}
\begin{figure}[htbp]
\centering 
\includegraphics[scale=0.9]{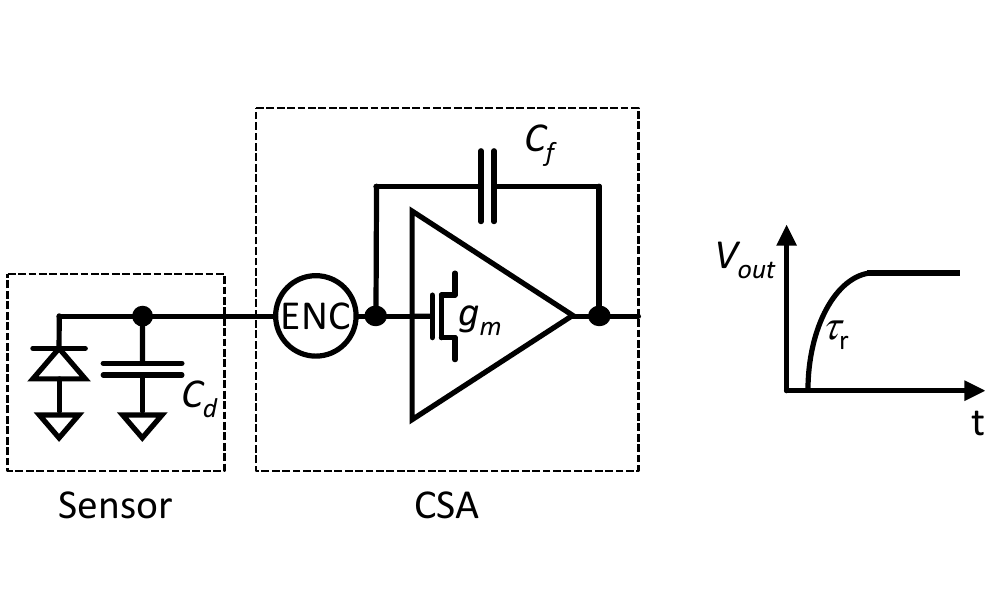}
\caption{Simplified equivalent circuit of a charge sensitive amplifier (CSA) connected to the n-side of a charge collecting diode of the sensor. The parasitic capacitance $C_\mathrm{d}$ at the input of the CSA is typically dominated by the junction capacitance of the charge collection electrode. Also shown are the input referred (series) noise source ENC and the output waveform with finite rise-time $\tau_\mathrm{r}$ in response to an ideal charge signal. The transconductance $g_\mathrm{m}$ of the CSA input device plays a central role in optimizing the noise and timing performance in the presence of the input capacitance $C_\mathrm{d}$.}
\label{fig:csa} 
\end{figure}

For the Large Hadron Collider (LHC) a luminosity upgrade from $\SI{e+34}{\per\square\centi\meter\per\second}$ to $\SI{e+35}{\per\square\centi\meter\per\second}$ (HL-LHC) is planned for 2026 \cite{HL-LHC}. This increase in luminosity requires an upgrade of the innermost pixel tracking layers since they will be exposed to ten times higher hit rates and radiation levels. To meet this challenge, a new generation of pixel sensors and read-out chips is being developed \cite{RD53}. The increase in hit rate demands smaller pixels to cope with the high track densities. At the same time thermal and mechanical constraints limit the allowed power consumption for the pixel layers. Therefore on-chip analog and digital signal processing of the hit data has to be carefully optimized to meet all performance specifications, including the power budget. The sensor capacitance influences the performance of the analog front-end as follows:

The circuit analysis of a typical charge sensitive amplifier (CSA, see fig.\ref{fig:csa}) shows that the pixel capacitance $C_\mathrm{d}$, which loads the input of the CSA, influences the analog performance of a detector system in terms of noise and timing behavior. Both the signal rise-time $\tau_\mathrm{r}$ and the series noise component $ENC_\mathrm{ser}$ (the equivalent input referred noise charge caused by the flicker noise and thermal noise of the input device \cite{Chang_Sansen}) are proportional to the capacitance $C_\mathrm{d}$. Also the charge collection efficiency $CCE$ depends on $C_\mathrm{d}$ \cite{Spieler}. These dependencies are expressed by the following relations: 
\begin{equation}
    \tau_\mathrm{r} \propto \frac{C_\mathrm{d}}{C_\mathrm{f} \cdot g_\mathrm{m}}
	\label{eqn:t_rise}
\end{equation}

\begin{equation}
    ENC_\mathrm{ser} \propto \frac{C_\mathrm{d}}{\sqrt{g_\mathrm{m}}}
	\label{eqn:ENC}
\end{equation}

\begin{equation}
    CCE \approx \frac{C_\mathrm{f}}{\frac{C_\mathrm{d}}{A_\mathrm{0}} + C_\mathrm{f}}
	\label{eqn:CCE}
\end{equation}

with the feedback capacitance $C_\mathrm{f}$, the input device transconductance $g_\mathrm{m}$, and the DC gain of the pre-amplifier $A_\mathrm{0}$. Equations \ref{eqn:t_rise} and \ref{eqn:ENC} show that for a given performance parameter a change of $C_\mathrm{d}$ can be compensated by adapting $g_\mathrm{m}$, which is proportional to the square root of the drain current $I_\mathrm{d}$ of the input device when operating in weak inversion, which is typical for analog front-end designs in pixel detectors. As changing $g_\mathrm{m}$ has a direct impact on the power consumption of the pre-amplifier, it is necessary to have a precise knowledge of the input capacitance $C_\mathrm{d}$ to be able to design an analog front-end with optimum performance and power efficiency. 

The capacitance of the sensor collection node depends on various parameters (implant geometry, isolation distances, doping density, etc.) which have to be optimized with respect to the charge collection efficiency - in particular after radiation damage - and the breakdown voltage which limits the maximum depletion bias. The discussion of these issues is beyond the scope of this publication. However, different collection node geometries will be briefly described in Chapter \ref{sec:sensor_design}. 

A typical sensor pixel with the size of $50 \times \SI{50}{\square\micro\meter}$ has a  capacitance in the order of a few tens of fF. Thus, a direct measurement, for example with an LCR meter, is challenging since the parasitic capacitance of the measurement connection  usually an order of magnitude larger than the capacitance to be measured. Nevertheless, in \cite{ATLAS_PIXEL_CAP} sensor pixel test structures have been characterized with a total measurement error of 5 fF using an LCR-meter, low parasitic capacitance probes, and a shielded probe station setup. However, one drawback of that approach is that such a setup requires dedicated sensor pixel test structures. In addition, only few pixels per structure (mostly only one) can be characterized that way. To overcome these limitations, we developed a dedicated integrated circuit (Pixcap65) that can be bump-bonded directly to any pixel sensor with a bump pitch of $50 \times \SI{50}{\square\micro\meter}$. In particular, the pixel matrix can be of any size and its design does not need dedicated features to enable the capacitance measurement.   

In section \ref{sec:measurement_method}, we will introduce the measurement method and in section \ref{sec:chip_implementation} we will explain how it is implemented in the Pixcap65 chip. The test sensor we used as a device under test is described in section \ref{sec:sensor_design} and finally the measurement results of the test pixel sensors bump-bonded to the Pixcap65 chip will be presented in section \ref{sec:measurement_results}.

\section{Measurement Method}
\label{sec:measurement_method}

\begin{figure}[htbp]
\centering 
\includegraphics[scale=1.2]{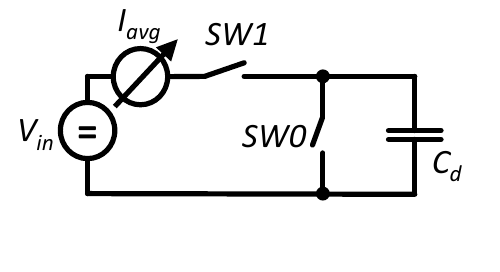}
\caption{Equivalent circuit diagram describing the charge-based capacitance measurement method. Periodic non-overlapping switching of SW1 and SW0 generates charge and discharge currents across the capacitance  $C_\mathrm{d}$ while only the average charge current that flows from the voltage source $V_\mathrm{in}$ is measured. The relation between the average charge current, the switching frequency and the DC voltage facilitates the extraction of the capacitance value.}
\label{fig:CBCM} 
\end{figure}

The pixel capacitance measurement implemented in the Pixcap65 chip makes use of the charge-based capacitance measurement method (CBCM)~\cite{CBCM}, which is schematically shown in Figure~\ref{fig:CBCM}. 

The basic measurement scheme consists of two switches SW0 and SW1, a constant voltage supply $V$, a DC current meter, and the capacitance to be measured. SW0 and SW1 are controlled by a non-overlapping clock sequence, which periodically charges the capacitance $C_\mathrm{d}$ to the voltage $V_\mathrm{in}$ (SW0 open, SW1 closed) and discharges to ground (SW0 closed, SW1 open). Since the charge and discharge currents have different paths, the average charge current $I_\mathrm{avg}$ can be measured separately with a DC current meter. With switching frequency $f$ and voltage $V_\mathrm{in}$, the capacitance value can be evaluated as:

\begin{equation}
    C_\mathrm{d} = \frac{Q}{V_\mathrm{in}} = \frac{\int_{0}^{T}i(t)\: \mathrm{d}t}{V_\mathrm{in}} = \dfrac{\frac{1}{T}\int_{0}^{T}i(t) \: \mathrm{d}t}{f \cdot V_\mathrm{in}} = \frac{I_\mathrm{avg}}{f \cdot V_\mathrm{in}}.
	\label{eqn:capacitance}
\end{equation}

Any additional parasitic current source (sensor leakage current, sub-threshold leakage of the switches) will cause an offset in the measurement. To avoid this systematic error, the capacitance is derived by measuring the current $I_\mathrm{avg}$ as a function of the switching frequency $f$ and applying a linear fit to the measured data. The capacitance is then given by the slope of $I_\mathrm{avg}(f)$ divided by the applied voltage $V_\mathrm{in}$ :

\begin{equation}
    C_\mathrm{d} = \frac{\mathrm{d}I_\mathrm{avg}(f)}{\mathrm{d}f} \cdot \frac{1}{V_\mathrm{in}}
	\label{eqn:cap_slope}
\end{equation}

Other systematic errors could be introduced by the on-resistance $R_\mathrm{on}$ of the switches. Combined with the capacitance, the on-resistance introduces a first-order low-pass filter with a time constant $\tau = R_\mathrm{on} \cdot C_\mathrm{d}$, which leads to finite charge and discharge times. To minimize this error, the maximum switching frequency (i.e., the minimum pulse width of the non-overlapping clock) has to be limited to allow charge and discharge voltages across $C_\mathrm{d}$ to settle to the required precision. On the other hand, the switching frequency must not be too low in order to allow for a reasonable precision of the current measurement (see section \ref{sec:measurement_results} for a more detailed discussion of the systematic errors).


\section{Implementation of the Pixcap65 Chip}
\label{sec:chip_implementation}

\begin{figure}[htbp]
\centering 
\includegraphics[scale=0.4]{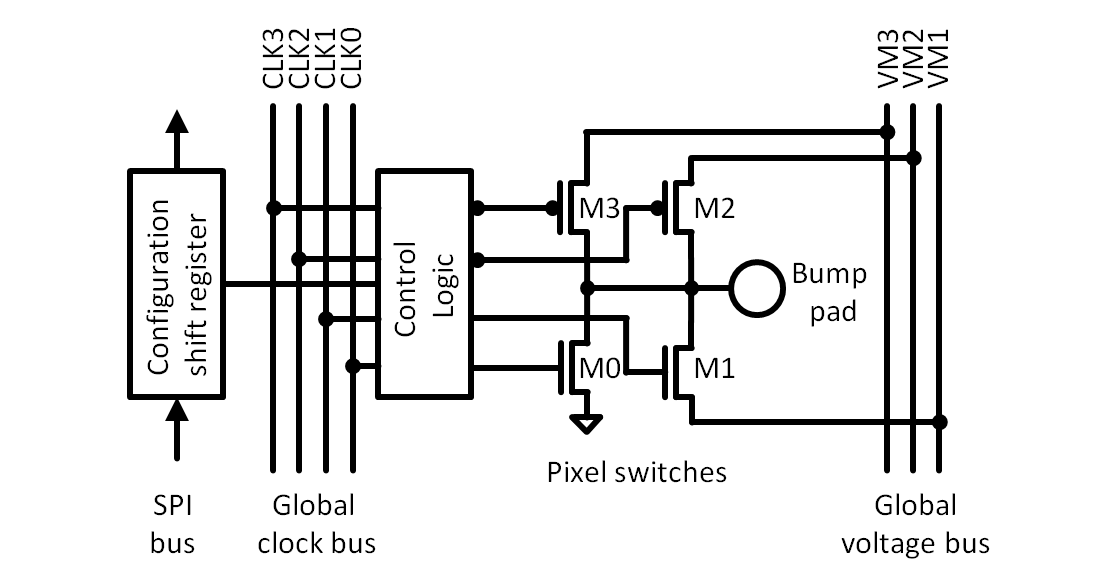}
\caption{Simplified circuit diagram of the Pixcap65 pixel cell. Each pixel has four MOS transistor switches which can be programmed to toggle the potential of the capacitance connected to the bump pad. The voltage levels are defined by the lines VM[3:1] and ground while the toggling sequence is controlled via the clock lines CLK[3:0].}
\label{fig:chip_pixel} 
\end{figure}

The Pixcap65 chip is an integrated circuit fabricated in a $\SI{65}{\nano\meter}$ CMOS technology. Its concept is similar to a previous chip implementation which was designed to be compatible with sensors with a pixel size of $50 \times \SI{250}{\square\micro\meter}$ \cite{PIXCAP_LF}.
The Pixcap65 chip has a $40 \times 40$ pixel matrix with a $50 \times \SI{50}{\square\micro\meter}$ pitch. This pitch makes the Pixcap65 chip compatible with planar and 3D sensors which are designed for the hybrid-pixel read-out chips of the HL-LHC upgrades of the ATLAS and CMS experiments at CERN. The pixel cell as shown in Figure \ref{fig:chip_pixel} consists of an array of switches, a control logic, and a bump pad to connect to a sensor pixel. The basic switch configuration to measure the total pixel capacitance only, as shown in Figure \ref{fig:CBCM}, is realized by a PMOS transistor M3, which charges the capacitance under test via the global voltage line VM3 and an NMOS transistor M0 to discharge the capacitance to ground. The transistors are switched via the global clock lines CLK3 and CLK0 respectively. The Pixcap65 chip has two additional sets of clock lines, voltage lines, and switch transistors (CLK1, VM1, M1, and CLK2, VM2, and M2) which enable more advanced measurements. The pixel control logic, which is programmed via a shift register, configures the connection between clock lines and switches for each pixel. This allows each pixel to either be toggled by a selective connection of the switch transistors to their corresponding clock net or to have a permanent connection to one of the potentials VM3, VM2, VM1 or ground. By choosing an appropriate pixel configuration and clock sequence, not only the total pixel capacitance can be measured but also displacement currents and thus the coupling between pixels can be resolved.

\begin{figure}[ht]
\centering 
\includegraphics[scale=0.4]{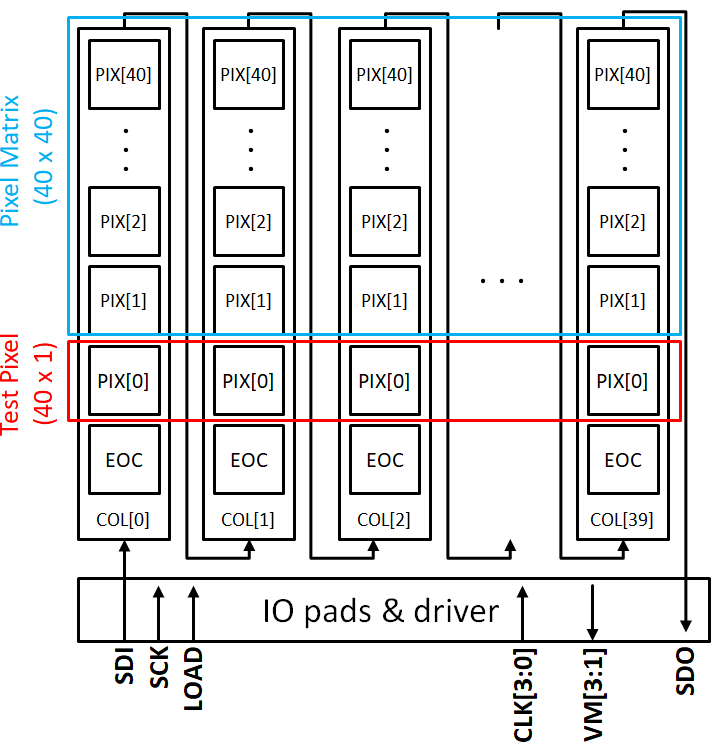}
\qquad
\includegraphics[scale=0.3]{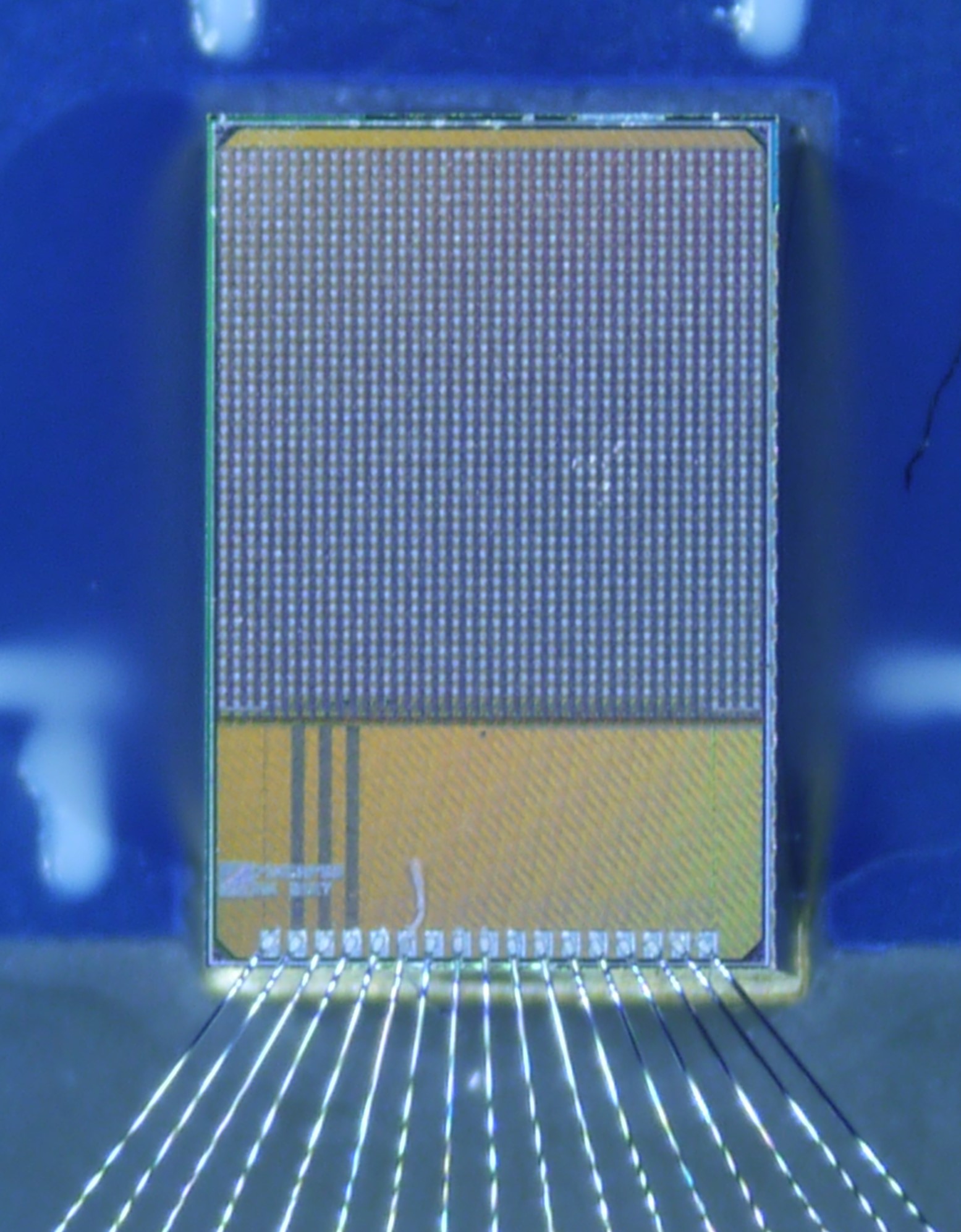}
\caption{Organization of the Pixcap65 chip (left) and photograph of a bare chip mounted on a test PCB (right). The chip has a size of $2040 \times \SI{3068}{\square\micro\meter}$. A matrix of $40 \times 40$ pixels has bump pads to be connected to a sensor. An additional row of pixels without or with unconnected bump pads is located at the bottom of the matrix whose cells are used for testing and calibration. An SPI bus (SDI, SCK, LOAD, and SDO) is used for chip/pixel configuration. Up to four externally generated clock sequences can be transmitted over the global clock lines CLK[3:0] while the voltage lines VM[3:1] connect to external source monitoring units which generate the constant charge and discharge potentials and measure the resulting currents.}
\label{fig:chip_organization} 
\end{figure}

 Electrically the pixels are grouped in columns as shown in the chip organization sketch in Figure \ref{fig:chip_organization}. Each pixel column has an end-of-column block (EOC) which controls the connection of the column-level voltage lines to the global voltage lines VM[3:1]. These additional (static) switches reduce loading of the voltage lines with leakage current from pixels in columns that are not active in a current measurement. Apart from the bump-bondable $40 \times 40$ pixel matrix, the chip has one additional pixel row at the bottom of the matrix. Its cells are connected to internal capacitors of various sizes which can be used for testing the chip without a sensor connected. A few of these extra pixels have an open output which facilitates an extraction of the parasitic capacitance of the bare switch circuit.

\section{Design of the Test Sensor Device}
\label{sec:sensor_design}

\begin{figure}[htbp]
\centering 
\includegraphics[width=.7\textwidth]{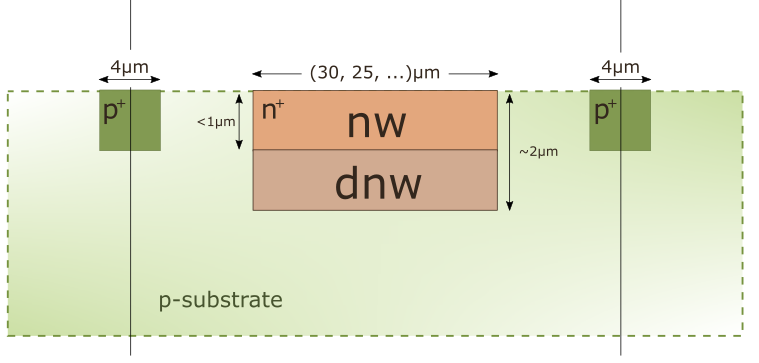}
\caption{Simplified cross section of a sensor pixel cell (metal layers are not shown). The n-type charge collection node is varied in depth (nw only or nw plus dnw implants) and implant width (30, 25, 20 and $\SI{15}{\micro\meter}$). The p$^+$ inter-pixel isolation ("p-stop") has a width of $\SI{4}{\micro\meter}$.}
\label{fig:crosssection} 
\end{figure}

\begin{figure}[ht]
\centering 
\includegraphics[width=.4\textwidth,]{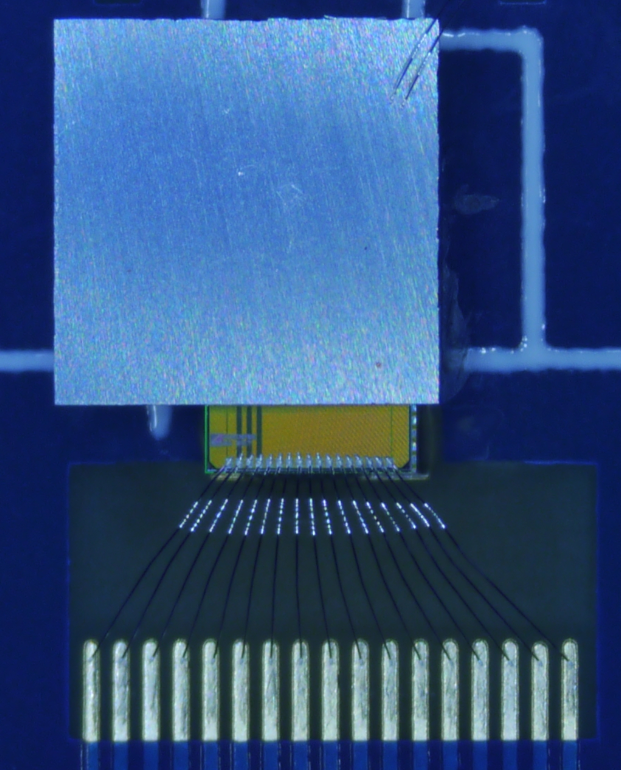}
\qquad
\includegraphics[scale=0.4]{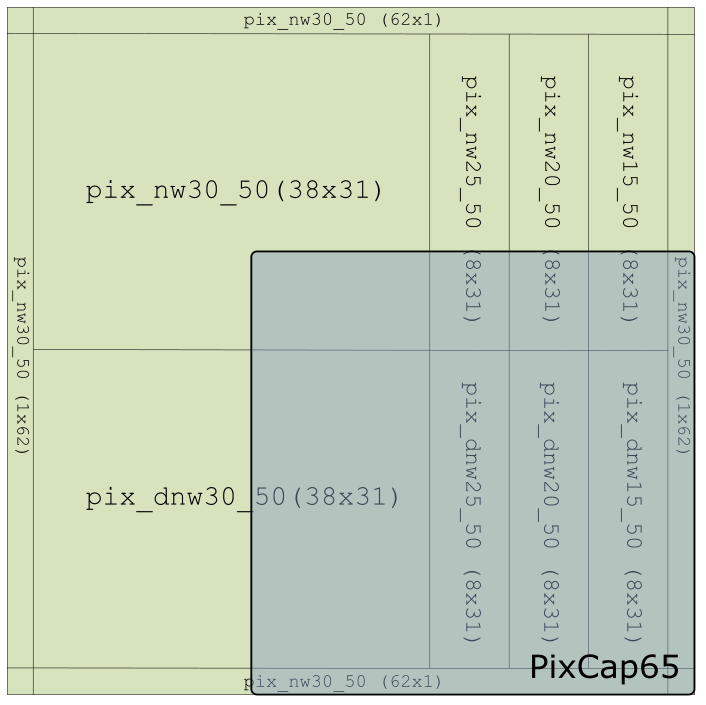}
\caption{Photograph of the Pixcap65 chip with attached test sensor on top (left) and organization of the pixel regions within the test sensor (right). The edge pixels are all of the same type (nw-implant only with $30 \times \SI{30}{\square\micro\meter}$ implant size). Pixcap65 chip and sensor are flip-chipped such that the active area of the chip is aligned with the lower right corner of the sensor. The available sensor has a size of 64 $\times$ 64 pixels and therefore extends over the Pixcap65 chip.}
\label{fig:chip/sens_photograph}
\end{figure}

To evaluate the performance of the Pixcap65 chip, we used a sensor prototype which was designed with variations in the sensor node geometry and thus varying pixel capacitances. This sensor is a planar n$^+$-on-n type and has $64 \times 64$ pixels with $50 \times \SI{50}{\square\micro\meter}$ pixel size and a p$^+$-implant grid ("p-stop") for inter-pixels isolation. Two basic parameters of the collection electrodes have variations in the design: The width of the n-implant and its depth. The aim of this design is to study how much the pixel capacitance can be reduced without affecting the charge collection performance which has been analysed in \cite{LF_TEST_SENSOR}.
Figure \ref{fig:crosssection} shows the simplified cross section of a sensor pixel cell. The p-implants define the pixel border and provide isolation between neighboring n-type charge collection nodes. While the distance between these p-type boundaries, and thus the pixel size, is kept constant the size and the depth of the n-implant is parameterized: The size of the n-implant is varied from $30 \times \SI{30}{\square\micro\meter}$ to $15 \times \SI{15}{\square\micro\meter}$ in four steps. Also the n-implant depth has two variants: a standard n-implant (nw), and a deeper n-implant using an n-well extended by a deep n-well (dnw). All possible eight variants are implemented on the pixels sensor. Since the Pixcap65 chip has $40 \times 40$ pixels it cannot cover the full sensor area of $64 \times 64$ pixels. However, the relative placement is chosen such that the Pixcap65 chip is connected to sensor pixels of all eight variants (see Figure \ref{fig:chip/sens_photograph}).

\section{Measurement Results}
\label{sec:measurement_results}
Measurements have been made with different assembly types: bare Pixcap65 chips with and without solder bumps, and chips assembled with the test sensor described in section \ref{sec:sensor_design}. Before the measurement results are presented, the main error sources of the measurement will be discussed. 

\subsection{Measurement Errors}
\label{sec:measurement_results_errors}
The accuracy of the sensor capacitance measurement is limited by two main contributions: the measurement error of the capacitance measurement and the dispersion of an offset correction. The first contribution depends on the error of the parameters of equation \ref{eqn:capacitance} (the voltage amplitude of the charge/discharge cycle, the current measurement, and the clock frequency), which define the error of the slope in relation \ref{eqn:cap_slope} and thus the measurement error of the total capacitance (i.e. the sensor capacitance plus the parasitic capacitance of the switch circuit). The second and dominating error contribution (as will be shown below) comes from the offset correction, which is necessary to suppress the parasitic capacitance from the switching circuit and its wiring. Inherent process variations of the CMOS technology lead to a dispersion of this parasitic capacitance, thereby limiting the accuracy of the absolute pixel capacitance measurement.

\paragraph{Clock frequency:}The clock oscillator which is used as a frequency reference has a precision of 150 ppm. The error on the frequency can therefore be neglected compared to the other sources.

\paragraph{Current measurement:}The measurement accuracy of the source-measurement-unit (Keithley 2410) is 0.029\% + $\SI{300}{\pico\ampere}$ at the $\SI{1}{\micro\ampere}$ range. This error is taken into account for the fit of a linear function to the current values measured as a function of the clock frequency.

\begin{figure}[htbp]
\centering 
\includegraphics[width=.7\textwidth]{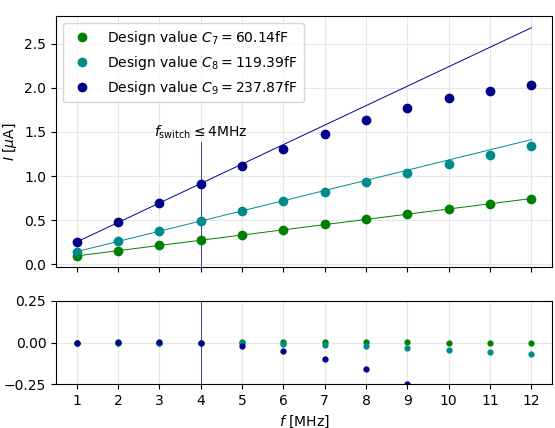}
\caption{Measurement of the displacement current as a function of the clock frequency. For large capacitances and high frequencies the deviation from a linear behavior becomes apparent when charge and discharge times approach the RC time constant given by the on-resistance of the switches and the measured capacitance.} 
\label{fig:rc_timeconstant1} 
\end{figure}

\begin{figure}[htbp]
\centering 

\includegraphics[width=0.7\textwidth]{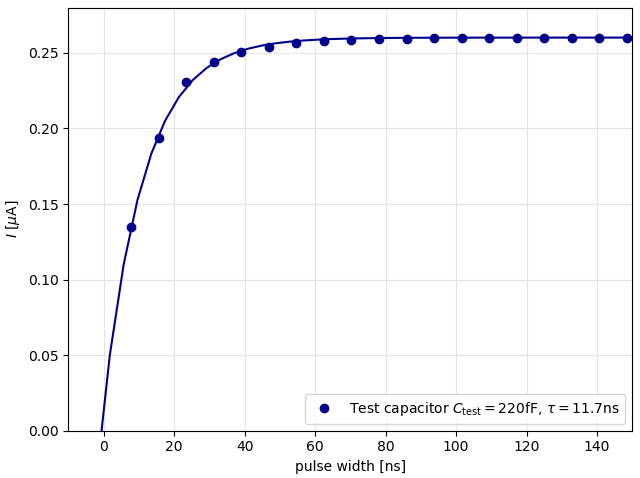}
\caption{Measurement of a large test capacitor ($\SI{220}{\femto\farad}$) at a constant frequency of $\SI{1}{\mega\hertz}$ as a function of the charge and discharge pulse width. The resulting time constant of $\SI{11.7}{\nano\second}$ yields $R_\mathrm{on} = \SI{53}{\kilo\ohm}$. }
\label{fig:rc_timeconstant2} 
\end{figure}

\paragraph{Voltage level:}The error of the voltage output of the source monitoring unit is 0.02\% + $\SI{600}{\micro\volt}$ at the $\SI{2}{\volt}$ output range, which is negligible. More important is the discussion of the systematic error resulting from the finite settling time $\tau$ given by the on-resistance $R_\mathrm{on}$ of the switches and the total capacitance $C_\mathrm{d}$. A too high clock frequency (i.e. a too narrow pulse width of the non-overlapping clocks) would reduce the voltage amplitude across the capacitance and thus underestimate the capacitance value (see Figure \ref{fig:rc_timeconstant1}). For example, to keep the contribution of the settling error below 1\% (0.1\%) the charge and discharge pulse width has to be larger than $5 \tau$ ($7 \tau$). To access the value of $R_\mathrm{on}$, the charge current into one of the large test capacitors ($C = \SI{220}{\femto\farad}$) has been measured as a function of the charge and discharge pulse width at a constant switching frequency of $\SI{1}{\mega\hertz}$ (Figure \ref{fig:rc_timeconstant2}). The resulting curve resembles the resistive charge-up of the capacitor and an exponential fit yields $\tau = \SI{11.7}{\nano\second}$ and thus $R_\mathrm{on} = \SI{53}{\kilo\ohm}$. With these numbers the settling error at a maximum switching frequency of $\SI{4}{\mega\hertz}$ is below 1\% (0.1\%) for capacitance values of up to $\SI{250}{\femto\farad}$ ($\SI{170}{\femto\farad}$).

\begin{figure}[htbp]
\centering 
\includegraphics[width=.7\textwidth]{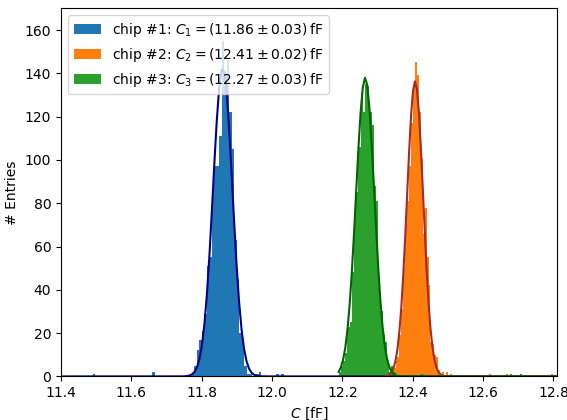}
\caption{Distribution of the pixel capacitance of the bare chips without bumps.}
\label{fig:histogram_no_bumps} 
\end{figure}

\begin{figure}[htbp]
\centering 
\includegraphics[width=0.7\textwidth]{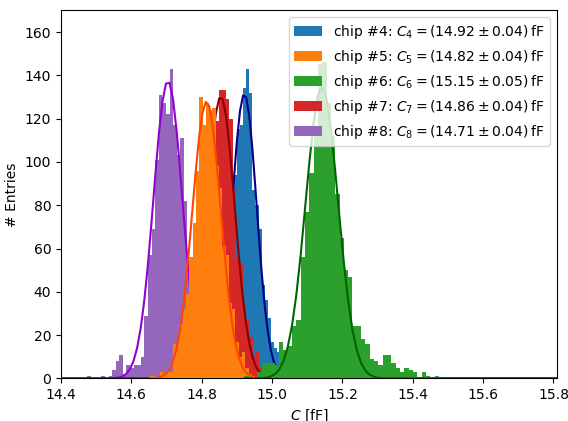}
\caption{Distribution of the pixel capacitance of the bare chips with bumps.}
\label{fig:histogram_bumps} 
\end{figure}

\paragraph{Process variations:} The systematic error introduced by the correction of the parasitic capacitance of the switch circuit is more difficult to access. The total capacitance measured with the described method is the sum of the capacitance under test (the sensor pixel capacitance) and the parasitic capacitance of the switch circuit itself. Process variations and finite matching of devices in CMOS processes lead to a dispersion of the circuit parameters that influence the parasitic capacitance. The result is a dispersion of the measurements across a pixel matrix as well as a dispersion of the pixel matrix average from chip to chip. Since the circuit parasitic capacitance can only be accessed with a Pixcap65 chip without sensor, a sensor measurement from a Pixcap65 chip with sensor suffers from a systematic error given by the dispersion of the parasitic capacitance. While the pixel-to-pixel dispersion within a chip is in the order of 0.05 fF and 0.03 fF for chips with and without bumps, respectively (see Fig. \ref{fig:histogram_no_bumps} and \ref{fig:histogram_bumps}), the spread of the average from chip to chip is larger. Since only a small number of samples (8 chips in total) was available, we estimated the error due to chip-to-chip dispersion to be about 0.3 fF (see also Table \ref{tab:rawresults}), which defines the calibration error of the parasitic switch capacitance (offset calibration) for the further sensor measurements.

In summary, it can be concluded that the accuracy of the sensor pixel capacitance measurement --- comparing all its contributions stated above --- is dominated by the inherent process variations of the CMOS technology resulting in an uncertainty of the parasitic capacitance calibration of about $\SI{\pm 0.3}{\femto\farad}$.

\subsection{Bare Chip}
\label{sec:bare_chip_measurements}
Bare chips (with and without solder bumps) have been measured to evaluate the parasitic capacitance of the measurement circuit in the pixel which would overestimate the sensor capacitance in the measurement if not corrected for. This parasitic capacitance of the circuit is mainly caused by the gate-drain capacitance of the switches, the wiring between switches and bump pad, and the bump pad itself. Measurements of the capacitance of bare pixel circuits (without solder bumps) show an average capacitance of $C_\mathrm{bare}=\SI{12.18 \pm 0.29}{\femto\farad}$ (average of all pixels, see Figure \ref{fig:histogram_no_bumps} and Table \ref{tab:rawresults}). The histogram in figure \ref{fig:histogram_bumps} shows the capacitance distributions of chips with attached solder bumps. The average capacitance of these samples is $C_\mathrm{bare+bump}=\SI{14.89 \pm 0.17}{\femto\farad}$. This common offset value is used for the correction of all following measurements of sensor assemblies. 

\begin{table}[htbp]
	\centering
	\caption{Measured capacitance values of the Pixcap65 samples. The switch and test capacitances are measured for all available samples including four assemblies with sensors.}
	\smallskip
	\begin{tabular}{|l|c|c|c|c|}
		\hline
		Capacitance type        & No. samples & No. cells/sample & $C_\textup{avg}$ [$\SI{}{\femto\farad}$]  &  $C_\textup{sigma}$ [\%] \\ \hline 
		Switch only             & 11          & 9                & $\SI{4.45 \pm 0.03}{}$                  &  0.66  \\
		Switch + test capacitor & 11          & 10               & $\SI{12.58 \pm 0.13}{}$                 &  1.01  \\
		Switch + pixel w/o bump & 3           & 1600             & $\SI{12.18 \pm 0.29}{}$                 &  2.34  \\
		Switch + pixel w/ bump  & 5           & 1600             & $\SI{14.89 \pm 0.17}{}$                 &  1.11  \\

		\hline
	\end{tabular}
	\label{tab:rawresults}
\end{table}

\subsection{Pixel Sensor Measurements}
\label{sec:chip_sensor_measurements}

Four samples of Pixcap65 chips flip-chipped to the test sensor described in section \ref{sec:sensor_design} are used for pixel capacitance measurements. The only difference in the samples is the thickness of the sensor substrate, which is available in $\SI{100}{\micro\meter}$ and $\SI{200}{\micro\meter}$. The applied bias voltage is $V_\mathrm{bias}=\SI{-80}{\volt}$ to ensure that the sensor volume is fully depleted. 

\begin{figure}[ht]
\centering 
\includegraphics[width=0.4\textwidth]{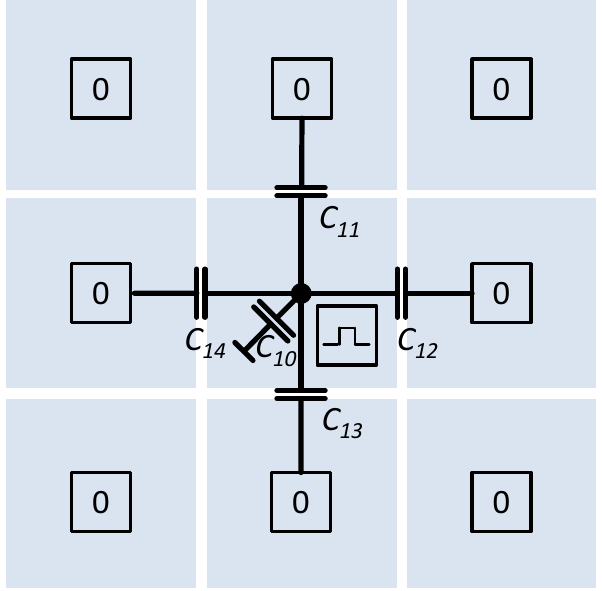}
\qquad
\includegraphics[width=0.4\textwidth]{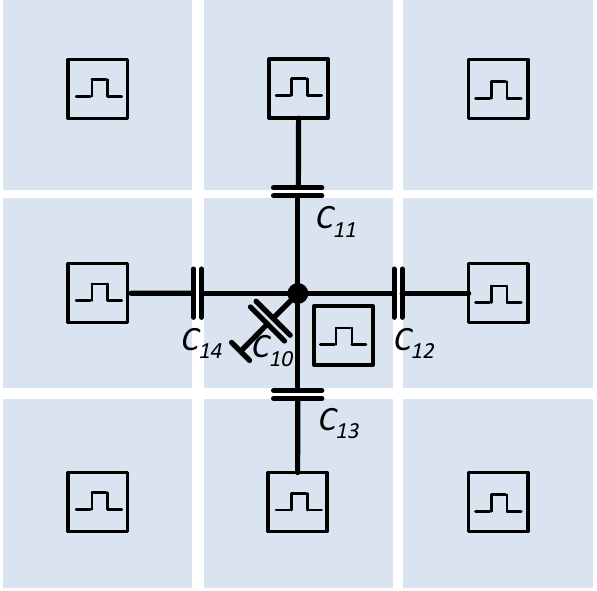}
\caption{Configuration for pixel total capacitance measurement (left) and configuration for suppression of the neighbor pixel capacitance (right). For the total pixel capacitance measurement, the central pixel is toggled and the resulting current measured while all neighboring pixel are connected to ground. In order to suppress the coupling to the neighboring pixels, they are toggled simultaneously with the central pixel. Thus, only $C_\mathrm{10}$ contributes to the measured charge current of the central pixel. For better readability, the diagonal coupling capacitances have been omitted.}
\label{fig:matric_configuration} 
\end{figure}

\paragraph{Total pixel capacitance:} For the measurement of the total pixel capacitance - which is the sum of the contributions from coupling to the neighbor pixels, the backside and the p-stop implant - the matrix is scanned with one pixel toggled and its current measured at a time while all other pixels are held static (connected to ground). The configuration is shown as a simplified sketch in Figure \ref{fig:matric_configuration} on the left. 

\paragraph{Inter-pixel capacitance:} To access the inter-pixel capacitance, the additional switch resources of the pixel circuit (see Figure \ref{fig:chip_pixel}) are used to toggle the neighbor pixels with the same timing and amplitude as the active pixel without their current added to the measured current. The configuration is shown in Figure \ref{fig:matric_configuration} on the right. Since this mode cancels the displacement current between the active pixel and its neighbors, their mutual capacitance ($C_\mathrm{11}, C_\mathrm{12}, C_\mathrm{13}$, and $C_\mathrm{14}$) is suppressed in the measurement. Thus, the measured capacitance in this mode - summarized as $C_\mathrm{10}$ - is the difference of the total capacitance and the sum of the coupling capacitances to the neighboring pixels.

\begin{figure}[htbp]
\centering 
\includegraphics[width=.8\textwidth]{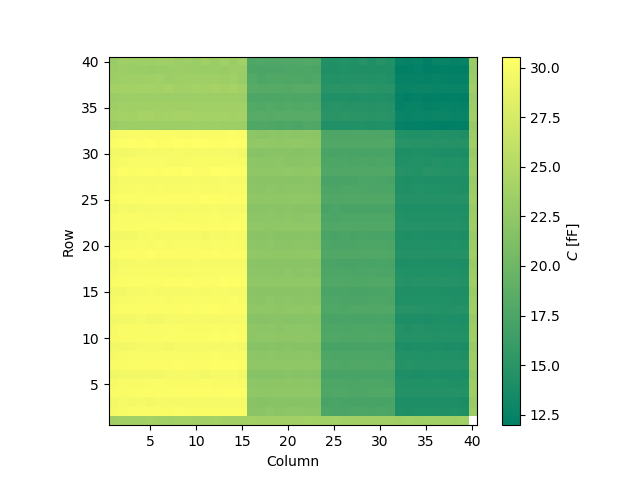}
\caption{2D map of the sensor pixel capacitance of sample 1.}
\label{fig:2d_sensor2} 
\end{figure}

In Figure \ref{fig:2d_sensor2} a 2D map of the measured total sensor pixel capacitances of sample 1 is shown. The different implantation areas can be clearly distinguished from each other. The pixel in the lower right corner is connected to a wire bond test pad, which leads to much higher capacitance. For the analysis this pixel was neglected. The sharp transitions indicate that the capacitance is dominated by the coupling between the n$^+$-implant of the pixel electrode and the p$^+$-implant of the isolation grid. If the pixel-to-pixel capacitance was the dominating contribution, the transitions between the differently implanted areas would be more blurred. The results of the total pixel capacitance and the inter-pixel capacitance measurements are summarized in Table \ref{tab:cresults}. The accuracy of the absolute capacitance values (total capacitance) is limited by the process variation discussed in section \ref{sec:measurement_results_errors} (estimated to be 0.3 fF). The error on the absolute inter-pixel capacitance is much smaller since it is a relative measurement where the error contribution of the parasitic capacitance correction cancels out. The standard deviations in Table~\ref{tab:cresults} are calculated from the distribution of four analyzed sensor assemblies.

\begin{table}[htbp]
	\centering
	\caption{Summary of the measured sensor pixel capacitances of different implantation types of the Pixcap65/test-sensor assemblies. Listed  are the total pixel capacitance $C_\mathrm{total}$ and the capacitance between two neighboring pixels $C_\mathrm{pix-pix}$ (mean values from four measured assemblies).}
	\smallskip
	\begin{tabular}{|l|c|c|c|c|}
		\hline
		Implant             & Columns & Rows   & $C_\mathrm{total}$ [$\SI{}{\femto\farad}$] &  $C_\mathrm{pix-pix}$ [$\SI{}{\femto\farad}$]\\ 
		\hline 
		\texttt{nw15\_50}   & 32-39  & 33-40 &   $\SI{12.42   \pm 0.15}{}$   &  $\SI{0.20   \pm 0.02}{}$ \\
		\texttt{nw20\_50}   & 24-31  & 33-40 &   $\SI{14.62   \pm 0.14}{}$   &  $\SI{0.26   \pm 0.02}{}$ \\
		\texttt{nw25\_50}   & 16-23  & 33-40 &   $\SI{17.84   \pm 0.13}{}$   &  $\SI{0.35   \pm 0.02}{}$ \\
		\texttt{nw30\_50}   & 1-15   & 33-40 &   $\SI{ 23.46  \pm 0.13}{}$   &  $\SI{0.47   \pm 0.03}{}$ \\
		\texttt{dnw15\_50}  & 32-39  & 2-32  &   $\SI{14.39   \pm 0.16}{}$   &  $\SI{0.27   \pm 0.02}{}$ \\
		\texttt{dnw20\_50}  & 24-31  & 2-32  &   $\SI{ 17.51  \pm 0.14}{}$   &  $\SI{0.38   \pm 0.03}{}$ \\
		\texttt{dnw25\_50}  & 16-23  & 2-32  &   $\SI{22.09   \pm 0.13}{}$   &  $\SI{0.54   \pm 0.03}{}$ \\
		\texttt{dnw30\_50}  & 1-15   & 2-32  &   $\SI{29.98   \pm 0.13}{}$   &  $\SI{0.80   \pm 0.03}{}$ \\
		\hline
	\end{tabular}
	\label{tab:cresults}
\end{table}

These measurements reveal that the sensor pixel capacitance increases with the width and depth of the pixel implantation, as expected. For identical implant dimensions, compared to the n-well implantation, the deep n-well flavour has a higher capacitance. No influence is seen due to the different thicknesses ($\SI{100}{\micro\meter}$ and $\SI{200}{\micro\meter}$) of the four analyzed sensor assemblies. This shows that the contribution of the pixel-to-back-plane capacitance is negligible since the sensor thickness is large compared to the lateral isolation distances between the implantations. We also note that the coupling between neighboring pixel implants is very small. The ratio between $C_\mathrm{pix-pix}$ and $C_\mathrm{total}$ is between 0.016 and 0.026. This implies that the by far dominating contribution to the pixel capacitance comes from the coupling to the p-stop implant. Other sensor types might behave differently, depending on the type of inter-pixel isolation, the use of a bias grid (the test sensor has no bias grid) and other layout details. 




\section{Summary}
In this paper we describe a pixel chip (Pixcap65) implementing the Charge-Based Capacitance Measurement (CBCM) method to precisely measure the capacitance of a pixel sensor. The chip can be flip-chip mounted to any pixel sensor with $\SI{50}{\micro\meter} \times \SI{50}{\micro\meter}$ pitch and facilitates a measurement of the total pixel and inter-pixel capacitances with sub-femtofarad precision using a simple desktop test setup. As a test device, the pixel capacitance of a planar n$^+$-on-n sensor with variations in the implant geometries was analyzed. It shows pixel capacitance values in the range of $\SI{12.42}{\femto\farad}$ to $\SI{29.08}{\femto\farad}$. The accuracy of the pixel capacitance measurement is estimated to be $\SI{0.3}{\femto\farad}$, limited by the dispersion of the parasitic capacitance due to the inherent process variations of the CMOS technology.


\acknowledgments
This project has received funding from the German BMBF under grant No. 05H15PDCA9. We thank the RD53 collaboration for the opportunity to have the Pixcap65 design being produced on a shared engineering run.




\begin{thebibliography}{99}

\bibitem{HL-LHC}
G. Apollinari et al., \emph{High Luminosity Large Hadron Collider HL-LHC}, \emph{arXiv:1705.08830}.

\bibitem{RD53}
RD53 Collaboration, \emph{Development of pixel readout integrated circuits for extreme rate and radiation}, \emph{CERN-LHCC-2013-008 ; LHCC-P-006}.

\bibitem{CBCM}
B. McGaughy, \emph{A Simple Method for On-Chip, Sub-Femto Farad Interconnect Capacitance Measurement}, \emph{IEEE Electronic Device Letters} {\bf 18} (1997).

\bibitem{Chang_Sansen}
Z.Y. Chang, W. Sansen, \emph{Low-noise Wide-band Amplifiers in Bipolar and CMOS Technologies},
Kluwer (1991).

\bibitem{Spieler}
H. Spieler, \emph{Semiconductor Detector Systems},
OUP Oxford (2005).

\bibitem{ATLAS_PIXEL_CAP}
G. Gorfine, M. Hoeferkamp, G. Santistevan, S. Seidel, \emph{Silicon Pixel Capacitance}, \emph{NIM A} {\bf 465} (201) 70-76.

\bibitem{LF_TEST_SENSOR}
Y. Dieter, M. Daas, J. Dingfelder, G. Giakoustidis, T. Hemperek, F. Hügging, J. Janssen, H. Krüger, D.-L. Pohl, M. Vogt, T. Wang, N. Wermes, \emph{Characterization of small-pixel passive CMOS sensors in 150 nm LFoundry technology using the RD53A readout chip}, \emph{NIM A} {\bf 972} (2020) 164130.

\bibitem{PIXCAP_LF}
M. Havrànek, Fabian Hügging, Hans Krüger, Norbert Wermes, \emph{Measurement of pixel sensor capacitances with sub-femtofarad precision}, \emph{NIM A} {\bf 714} (2013) 83-89.








\end{thebibliography}
\end{document}